# Detecting COVID-19 from Chest Computed Tomography Scans using AI-Driven Android Application

Aryan Verma, Sagar B. Amin, Muhammad Naeem, Monjoy Saha*

*Abstract*—The COVID-19 (coronavirus disease 2019) pandemic affected more than 186 million people with over 4 million deaths worldwide by June 2021. The magnitude of which has strained global healthcare systems. Chest Computed Tomography (CT) scans have a potential role in the diagnosis and prognostication of COVID-19. Designing a diagnostic system which is cost-efficient and convenient to operate on resource-constrained devices like mobile phones would enhance the clinical usage of chest CT scans and provide swift, mobile, and accessible diagnostic capabilities. This work proposes developing a novel Android application that detects COVID-19 infection from chest CT scans using a highly efficient and accurate deep learning algorithm. It further creates an attention heatmap, augmented on the segmented lung parenchyma region in the CT scans through an algorithm developed as a part of this work, which shows the regions of infection in the lungs. We propose a selection approach combined with multi-threading for a faster generation of heatmaps on Android Device, which reduces the processing time by about 93%. The neural network trained to detect COVID-19 in this work is tested with F1 score and accuracy, both of 99.58% and sensitivity of 99.69%, which is better than most of the results in the domain of COVID diagnosis from CT scans. This work will be beneficial in high volume practices and help doctors triage patients in the early diagnosis of the COVID-19 quickly and efficiently.

## I. INTRODUCTION

After the outbreak in China in December 2019, the World Health Organization (WHO) identified Severe Acute Respiratory Syndrome CoronaVirus-2 (SARS-CoV-2) as a new type of coronavirus. COVID-19 is the disease caused by SARS-CoV-2, which primarily affects the respiratory system.

The coronavirus 2019 breakout was declared a public health emergency at international level by the World Health Organization on 30 January 2020. It was given a pandemic status on 11 March 2020. Economies were ruined by the pandemic and it caused unrivalled challenges to healthcare and food system across the globe. The pandemic has overwhelmed health care systems. As a result, diagnosing and treating other diseases have been postponed. After multiple waves of COVID-19, health care workers and society in general have become exhausted.

The real-time reverse transcription-polymerase chain reaction (RT-PCR) test is the standard and used for detecting the presence of COVID-19 in an individual [1]. Due to the high false-negative rates, long turn around times, and shortage of RT-PCR kits, chest CT scans were found to be an effective and fast alternative to diagnosing COVID-19 [2]. CT scans combine series of X-ray images taken from different angles around the chest which are then post processed by computer to create detailed cross-sectional images Chest CT scanning is valuable in the diagnosis of COVID-19 disease.

In some cases, the RT-PCR gave negative results and is highly operator dependent, but CT scans confirmed the diagnosis of COVID-19 [3]. Overall due to the high specificity and fast diagnosis, chest CT findings can be a better option than RT-PCR [4]. On chest CT, COVID-19–associated pneumonia usually has a pattern of ground-glass opacification in a peripheral and lower lobe distribution in the lungs[5][6]. This common imaging pattern on CT was used as an aid to observe the COVID-19 in lungs through deep learning[7][8]. In this work, the neural network was trained to detect this finding with very high accuracy and specificity, and fewer parameters, as the model is to be ported on a mobile device.

Most of the devices consisting of the Android Operating System (OS) consist of mobile phones and tablets. Given the dire situation caused by COVID-19, a portable, swift, and completely automated aid for diagnosing the disease on CT scans would be beneficial. The CovCT app can provide this service on a portable android device. Unlike time-consuming testing methods such as RT-PCR, CovCT app would provide an optimal method of triaging patients with COVID-19. Also, the CovCT android application is capable of generating heatmaps to further illustrate affected regions in the lung.

Aryan Verma is with Department of Computer Science and Engineering, National Institute of Technology, Hamirpur, HP-177005 India (e-mail: aryanverma19oct@gmail.com).
Sagar B. Amin is with Department of Radiology and Imaging Sciences, Emory University School of Medicine, Atlanta, GA-30322, USA (e-mail: sagar.b.amin@emory.edu)
Muhammad Naeem is with Department of Radiology and Imaging Sciences, Emory university School of Medicine, Atlanta, GA-30322, USA (e-mail: muhammad.naeem@emory.edu)
Monjoy Saha is with Department of Biomedical Informatics, Emory University School of Medicine, Atlanta, GA-30322, USA (e-mail: monjoybme@gmail.com) (*Corresponding Author: Monjoy Saha)

## II. Related Work

There were many methods developed with the use of Artificial intelligence for detecting Covid-19 infection from chest CT findings. Some of the most important and recent advancements have been stated in Table I. Some of them used minimal approaches such as hierarchical and spatial models to detect COVID-19 relates features from chest CT [9]. Various deep learning approaches were employed in the direction of better accuracy for COVID-19 related findings from CT Images [10][11][12][13][14][15][16]. Gianchandani et al. proposed an ensembling method for COVID-19 diagnosis from chest X-rays through an ensemble of deep transfer learning models for better performance [17]. Hasan et al. proposed the Coronavirus Recognition Network (CVR-Net) in their work, which uses radiography images to detect COVID-19. The results from this showed an average accuracy of 78%. [18]

Accuracies in models trained with transfer learning were also observed to be very good. Brunese et al. reported an average accuracy of 97% in their work. They used a pre-trained VGG-16 model and performed transfer learning on the model for automatic detection of COVID-19 using chest X-Ray images [19]. Jaiswal et al. performed transfer learning of the DenseNet201 Model for the classification of the COVID-19 infected patients. It extracted features by using its learned weights on the ImageNet dataset [20]. T. Anwar et al. used EfficientNet B4 to distinguish between COVID and normal CT-scan images with a 0.90 F1 score. [21]

Many approaches customized deep learning architectures for better detection. Ozturk et al. presented DarkCovidNet which automatically detected COVID-19 using chest X-ray images. The classification accuracies obtained from this model were 98.08% for binary cases [22]. In another work, Mukherjee et al. developed a custom architecture of CNN which had nine layers for detecting COVID-19 cases. For the training of the model, they used X-Rays, and CT scans. The network achieved an overall accuracy of 96.28%, which was better than most of the CNN-based models [23]. In the Al-Karawi et al., Gabor filters extracted different texture features from CT images. Then these features were utilized for training support vector machines, which were further employed for classifying the Covid-19 cases. This approach got an average accuracy of 95.37% and a sensitivity of 95.99% [24]. N. Palaru et al. proposed Anam-Net, which is a CNN architecture based on depth embeddings. It detected irregularity in COVID-19 chest CT images. The Anam-Net architecture was lightweight and can be used for inference generation in mobile or resource constraint platforms [25]. H. Alshazly et al. studied various deep network architectures such as SqueezeNet, Inception, ResNet, Xception, ShuffleNet, and DenseNet and proposed a transfer learning strategy to achieve the best performance. As a result, ResNet101 achieved the average accuracy of 99.4% which is better than others and had an average sensitivity of 99.1% [26]. N. Basantwani et al. used transfer learning on an Inception-V3 model and ported it to an android application with an accuracy of 94% [27]. Most of the models were observed to be either having less accuracy or a very large number of parameters for the model to be ported to a mobile device. Some models were customized for mobile devices, but our approach is entirely different from them and outperforms them for a perfect blend of accuracy, parameters, and specificity.

## TABLE I

The Study of Various Existing Machine Learning techniques for COVID-19 Detection from chest CT scans and their corresponding Results. All of these studies are aimed at Covid and No-Covid Classification of CT images.

| S. No. | Method | Results | Reference |
|---|---|---|---|
| 1 | Transfer learning on Inception Recurrent Residual Neural Network | Accuracy – 98.78% | [28] |
| 2 | Construction of AI model using Transfer learning on ShuffleNet V2 | The area under the curve (AUC), sensitivity of model and specificity of model were 0.9689, 90.52% and 91.58% respectively. | [29] |
| 3 | Feature extraction done by DenseNet121 and bagging classifier trained on top of these | Accuracy – 99 ± 0.9% | [30] |
| 4 | Lesion-attention deep neural networks, using pretrained network weights including VGG16, ResNet18, and ResNet50 | with 0.94 of the AUC score | [31] |
| 5 | Comprehensive System using ResNet 50 | sensitivity, specificity, and the AUC score were 94%, 98%, and 0.9940 respectively | [32] |
| 6 | Network based on regression of multi view point and 3-Dimensional U-Net | accuracy and sensitivity of 94% and 100% | [33] |
| 7 | Transfer learning on ResNet18 | AUC score – 0.9965 | [34] |

## III. Overview of Approach

This work is motivated to design a portable and accurate COVID-19 diagnosis system working on an Android application with a Deep Learning algorithm to analyze the chest

CT scans for presence of COVID-19 and further mark the COVID-19 presence in the segmented lungs' region through an attention map. The utmost focus lies in designing the deep neural network with high accuracy and fewer parameters that can be deployed to the Android Operating System (OS) using minimal memory, unlike many other works that render the trained models in hundreds of Megabytes (Mb).

class from the trained neural network due to its accurate and soft map formations. The algorithm is ported to android using Java programming language. The ScoreCAM algorithm takes a lot of time (9-10 Minutes) to form the heatmap, so a selective approach on activation maps and Multi-Threading environment in android, as explained in this work, is applied before passing the neural network activations. This approach reduces the time

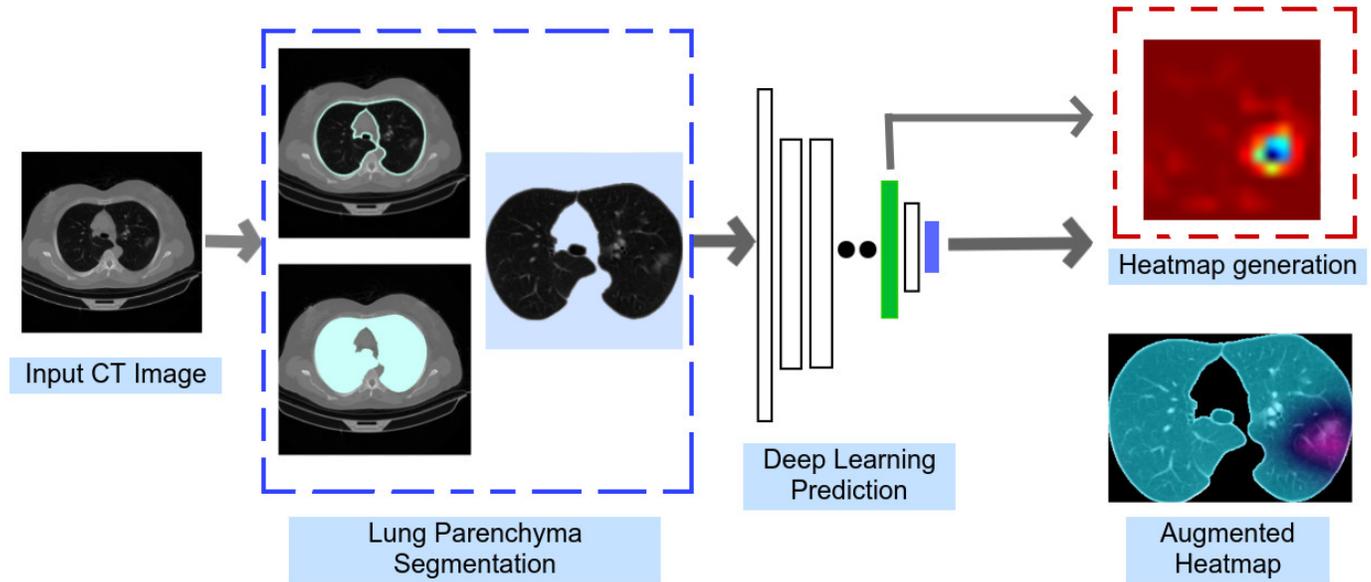

Fig. 1. An overall representation of the approach is shown. The blue dashed line is for the steps of lung parenchyma segmentation algorithm proposed as a part of this work. The red dashed line represents the algorithm for making and augmentation of heatmap. The green layer in the neural network is the last Convolutional layer of the deep learning algorithm proposed in the work and blue layer is the softmax layer of the deep neural network.

For an easy view and file storage in an android device, the chest CT scans are normalized and converted to Portable Network Graphics (PNG) Format. Input received in the form of a PNG image is fed to the lung parenchyma segmentation algorithm developed in this work using computer vision. The algorithm proposed for lung parenchyma segmentation is robust and involves very light processing on a mobile device. This segmentation algorithm first forms the contour around the detected regions of lung parenchyma and then segment the image for the lung region as shown in the blue dashed box region of Fig.1. This segmented region is enlarged and displayed on the screen for a better view of lung parenchyma.

The image is then passed to the deep learning model which is an EfficientNetB0 model fine tuned with further explained transfer learning and custom metrics to accept 512 X 512 images and outputs the class probabilities for covid detection. If the prediction result is covid positive, it is allowed to move to the next stage, the heatmap generation. The heatmap is further augmented over segmented lungs, as shown in the red dashed box region of Fig.1. this stage requires the output of the last convolutional layer (indicated by the green layer in Fig.1) along with the model predictions to generate the heatmap.

After testing a few algorithms to form heatmaps, the Score-CAM algorithm is used to generate the heatmap of the covid taken to develop the heatmap to 50-60 seconds, which is further dependent on the android device's performance. Finally, the heatmap is augmented over the chest CT scan and masked to show the segmented lungs region for a better inference. The heatmap can be adjusted for its hue values and gradient values using image processing.

## IV. MATERIAL AND METHODS

### A. Dataset

In this work, for the training and testing purpose of the neural network developed, the COVID-CTset dataset is used [35]. This dataset comprises of CT scans of total 377 patients consisting of 63849 CT scan images. 15589 CT scan images belong to 95 patients affected with COVID-19 and 48260 CT scan images belonging to 282 non-COVID patients. The CT scans were gathered from Negin medical center, Sari, Iran. The original files in the dataset were in Tagged Image File Format (TIFF) format containing 16-bit grayscale data and did not include the patients' private information.

Android devices or regular monitors do not visualize the 16-bit grayscale TIFF images. There is a separate algorithm for visualizing these TIFF files, given by the dataset authors. So, to make it accessible and simply visualize images, TIFF files were converted to 8-bit PNG images through the normalization process in this work. Converting TIFF image files to PNG images gave a better view and analysis of these images on android platform. The tonal values of the TIFF image pixels range from 0 to over 5000, So if each image is scaled based on the maximum tonal value, it can cause data loss and reduce the performance of network. For tackling this issue, we trained the neural network to detect the COVID-19 related findings not from the TIFF image files but the PNG images. Which gave more satisfactory results as any image uploaded to the android platform in PNG form was easy to visualize and process.

To search for the best hyper parameters of the neural network and optimizer, five-fold cross validation was used. For this purpose, training and testing data was taken in five folds, provided by the author of the dataset. In each fold 20% of the data was used for testing. The model was trained on each fold of the data and tested on the corresponding test data fold.

After searching the hyper-parameters for the best accuracy of the neural network, the whole dataset was rearranged by combining the five folds of data that were earlier distributed into train, test, and validation data. The distribution statics of train, validation and test data is shown in Table II. Train and validation data was used to train the final model, and after training, the model was evaluated on the test data.

## TABLE II

The Train Dataset was Balanced for a Better Learning of the Neural Network. The Test Dataset was also Balanced for a Proper Testing of the Trained Model.

| Dataset | COVID-19 images | Normal Images |
|---------|-----------------|---------------|
| Train | 9128 | 9618 |
| Validation | 2282 | 39262 |
| Test | 2282 | 2250 |

### B. Normalization of TIFF Images to PNG

The TIFF files contain pixel values ranging from 0 to 5000, which are rendered as a black image on android mobile phones. This makes the identification and processing complex. To solve this, a normalization process is applied to each TIFF image file in the dataset, and all images are converted into normalized PNG form, which is easily visible and processed on android OS.

The process of normalization used here is the Min-Max Normalization applied with the help of OpenCV Library. Min-max normalization is applied, being the most common way to normalize the data. For every location in the TIFF image, the tonal value of the pixel is normalized according to the formula shown in Fig.5. For this work, the maximum and minimum of the second function (g) are 65535 and 0, respectively. The wide range is chosen to enhance the contrast of the images before converting them to PNG. If the maximum of the second function is too small, all the images will appear black due to less contrast and tonal values. Hence, through normalization, the minimum value in the TIFF image gets transformed into 0, the maximum value gets changed to 65535, and every other value gets changed into a number between 0 and 65535, according to the formula

$$v' = \frac{v - min_f}{max_f - min_f}(max_g - min_g) + min_g$$

[36]

Where $f$ is the input function and $g$ is the output function. Here, $v$ is the original value of pixel and $v'$ is the normalized value.

Now, the normalization of the TIFF image produces an output with values from 0 to 65535, so it is again divided by 255 to convert it into values between 0 and 255, as shown in Fig.2. While saving the image, the decimal data values from the output get rounded off to the nearest integer. This gives us the image intensity values suitable to be represented on normal monitors and android devices. Hence, the data can be saved as PNG images. This is done with whole dataset images, and TIFF files are converted to PNG files before feeding the data to the pipelines, used for training the model. This process not only is useful for the training, but the display of the files is easier in PNG form in Android OS dependent devices.

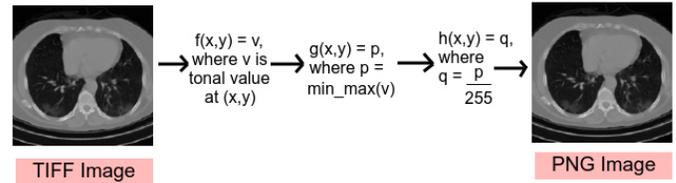

Fig. 2. The conversion of TIFF image to PNG image using Min-Max normalization and division.

### C. Lung Parenchyma Segmentation

The portion of the lungs which is involved in the transfer of gases is known as lung parenchyma and includes alveoli, alveolar ducts, bronchioles, and other essential tissues. The CT scan images the entire chest which includes the esophagus, trachea, heart, lungs, diaphragm, thymus gland, aorta, spine, nerves, veins, and arteries. Also, the approach proposed in this work generates a heatmap only for detecting the COVID-19 related finding in the lungs. Hence, the region concerned with the diagnosis of COVID-19 is the lungs, and the rest organs must be segmented from the CT scan for a proper view of any COVID-19 related findings. Figure 3 shows an annotated image of the chest CT scan. For this complex orientation to be resolved into a simpler view of lungs, this algorithm is proposed.

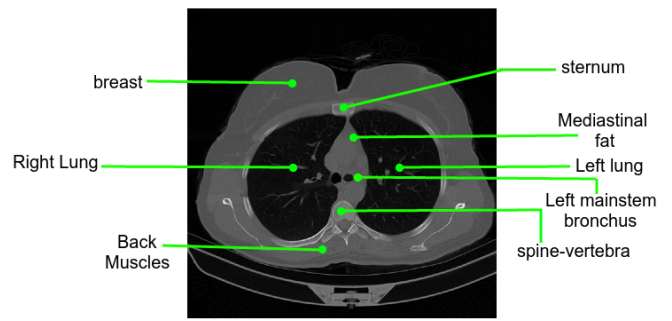

Fig. 3. An annotated CT image slice from the dataset. It shows the complex orientation and view of a regular CT image which makes the diagnosis difficult owing to stressed visibility of lungs parenchyma.

For a proper diagnosis, this segmentation algorithm takes out the region of the lungs parenchyma for a better view and analysis of the chest CT Image in an android device and magnifies it to image dimensions by series of image processing operations. This segmented mask is used to augment the generated heatmap to show the COVID-19 affected regions in the lung parenchyma. This algorithm is less complex and time-efficient to run on an android device as compared to other approaches which use complex algorithms. [37-42]

The chest CT scan uploaded on the CovCT Application is read into the memory in the form of a Bitmap with channel ordering alpha, red, green, and blue. This Bitmap is converted into an OpenCV Mat using the Utils package of Android. The mat image is still in ARGB form, is converted to grayscale. This conversion is necessary for further operations taking place. After converting the ARGB image to grayscale, the global thresholding algorithm, Otsu [43], is applied to get a binary image as shown in Figure 4. A binary image has either white or black pixels, which determine the foreground and background respectively.

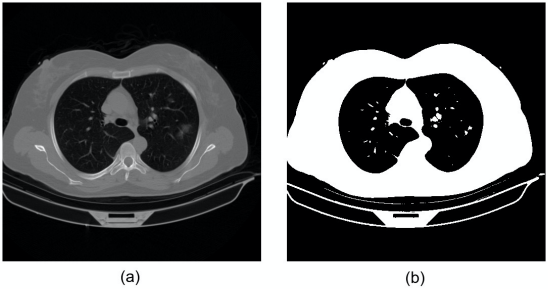

Fig.4. Image labelled (a) is the original CT image and (b) is the corresponding Otsu thresholding applied to the (a) image.

As the image is normalized during its conversion to PNG, the area representing lung parenchyma, diaphragm, and small other areas are highlighted in the images. Due to which the histogram of the image shows two clearly expressed peaks. The value which minimizes the weighted variance of these two clusters of the histogram is taken as the threshold value.

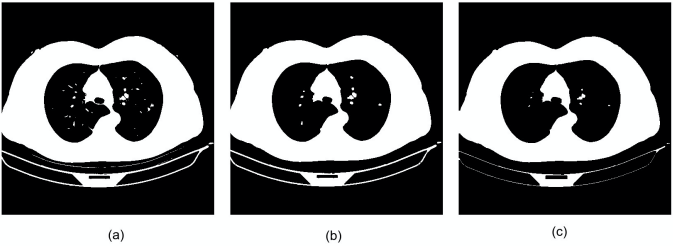

Fig. 5. Image labelled (a) is the otsu binarisation result. Image (b) is the result when morphological opening operation is performed on (a) image to remove impurities inside it. Image (c) is the result after dilating the (b) image, which have less number of holes inside lung parenchyma region.

The thresholded binary image is passed through some morphological image processing operations to remove the impurities from the foreground objects. Morphological opening operation with a 3x3 kernel is performed on the binary output of the thresholding algorithm, which first dilates the image for removing the holes and impurities inside the foreground mask and then erodes it for keeping the size of the foreground the same. The result is displayed in Fig. 5 (b). The resulting foreground mask is filled with holes but does not cover the lung parenchyma boundary, as it remains shorter than it. To let the foreground mask cover the entire region, it is again dilated two times as shown in (c) part of Figure 5.

The resulting image is a binary mask with foreground pixels representing the lung parenchyma, diaphragm, and maybe 1-2 small findings. Now to segment the lung parenchyma from the binary mask, contour detection is applied. A binary mask is provided as an input to the contour finding function. The function finds the complete contours of the foreground regions in the binary image along with the image's border and makes a tree-like hierarchy. The output of the function is the list of all detected regions in the foreground.

Fig. 6. Image labelled (a) is the original CT image on which sorted

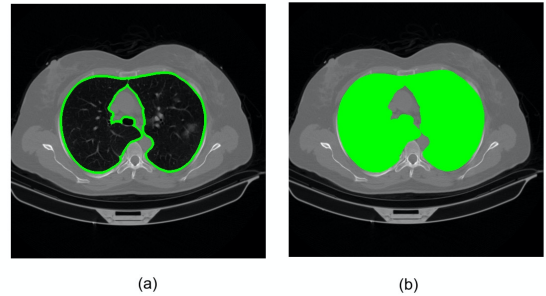

contour is drawn from selection construct. Image (b) is the corresponding result after the contour is filled and is used for final mask generation.

This list is iterated for finding the lung contours using a selection construct that checks every contour for its area and sorts out contour with an area less than 512x512 and more than 100x100. This iteration process selects the lung parenchyma contour from the list as shown in Fig.6. The contour is further used to generate a binary mask used for segmenting the lung parenchyma region from the CT image using bitwise and operation. The segmented lung parenchyma from the CT scan is finally enlarged using a rectangle approximation from the detected contour and displayed. This enlarged view shows the lung parenchyma more clearly and makes the diagnosis of COVID-19 easier for the radiologists.

### D. Deep learning Model Architecture

This section explains the different approaches used in this work to find the optimal architecture for the deep learning model to be trained for detecting COVID and No-COVID from the CT scans. The hyperparameters in the neural network are wisely calculated by hit and trial along with some bilinear interpolation. The size and interim top layers of the model are also decided by training and testing the model on five folds of data and cross-validating it. Finally, all the determined and calculated parameters are used to train the model with all the data present in five folds.

After going through previous works in the domain of Covid detection from chest CT scans, results from various works are analyzed for the selection of the best deep learning architecture that must be used as a base model for transfer learning tasks. Resnet50V2 and EfficientNetB0 deep neural architectures are observed to be best for the classification task of CT scans in to COVID and No-COVID images.

After changing the input layer in both the models to accept the grayscale 512x512 CT scan image, both the models are trained and evaluated on the five folds of data without any hyper-parameter optimization. This process is done to select one out of these two architectures for final training and optimization. Observing the results in Table III, the EfficientNetB0 outperforms the Resnet50V2 model in four out of five folds of the data. Further, the number of parameters in EfficientNet B0 is 5.3 Million (Approx.), which is significantly less than Resnet50 V2 with 23 Million params (Approx.). Being less in number of parameters and more accurate, EfficientnetB0 is very suitable for being ported to mobile devices.

| Model | Fold | Training Accuracy | Testing Accuracy |
|---|---|---|---|
| Resnet50 V2 | Fold 1 | 96.72 | 95.88 |
| Efficientnet B0 | Fold 1 | 97.13 | 96.21 |
| Resnet50 V2 | Fold 2 | 96.32 | 95.91 |
| Efficientnet B0 | Fold 2 | 97.05 | 96.00 |
| Resnet50 V2 | **Fold 3** | **97.19** | **96.20** |
| Efficientnet B0 | **Fold 3** | **96.73** | **96.07** |
| Resnet50 V2 | Fold 4 | 97.71 | 96.23 |
| Efficientnet B0 | Fold 4 | 97.98 | 97.19 |
| Resnet50 V2 | Fold 5 | 96.85 | 95.93 |
| Efficientnet B0 | Fold 5 | 97.82 | 97.01 |

Hence, The deep learning model which is used for classification of CT scans for COVID or no-COVID is built on the top of EfficientNetB0 architecture. EfficientNet is a family of convolutional neural networks (CNN) which is formed after uniformly scaling the network dimensions with specific coefficients.

The input of the classification model is of shape 512 x 512 x 1, which is a single-channel image. The EfficientnetB0 model takes a image with three channels, and our input data is a single channel image, so we make an input layer with our defined shape, i.e., 512 x 512, and add the base as efficientnetB0 layers without including the top layers. The top layer is excluded in order to configure model output for two classes.

After adding the Input layer and EfficientNetB0, a global average pooling layer is added on the top of the base of the classification model. One feature map is generated for each corresponding category. Instead of adding fully connected layers after the formation of the feature maps, the average of each of these feature maps is taken, and the vector which comes as result is fed directly into the dense layer with a dropout layer in between, as shown in Fig. 8. Now, to select proper dropout from 0 to 1, the EfficientNet model is again passed through a test in which the same model is trained with different dropouts on the data, and the dropout of 0.3 is found to be optimum. For preventing the overfitting of the model, a dropout is placed and is experimentally verified to give better results in this case.

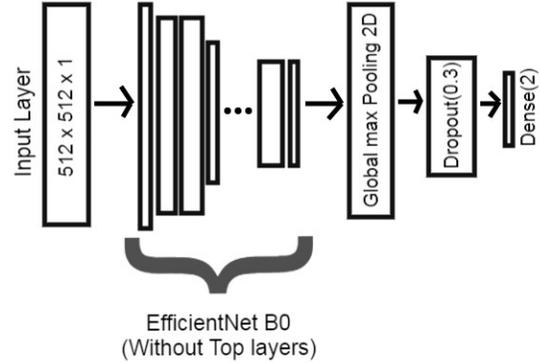

Fig. 7. Decided Architecture of the neural network, to be trained for detection of COVID-19 from PNG images of CT scans. The complete architecture of EfficientNetB0 is explained in supplementary section.

The dense layer contains the softmax activation function, which further renders the final output of the classification model to two probabilities, the first value gives the probability of covid-19 presence, and second value gives the probability of no-covid-19.

### E. Model Training and Testing

The code for the model's training has been designed using Keras, a high-level API for deep learning tasks. It uses TensorFlow as its backend. Data from the dataset is organized into three data generators which feed the data to the training algorithm of the model. The data from these generators are sent to the model in the form of batches. Each batch contains a fixed number of input data samples. The batch size used for train and validation generators is 10 each, and for test generator is 20.

The model is created using Keras according to the specified architecture in section IV.D, and weights are initialized for the base layers as the Imagenet weights. Imagenet weights are weights of the EfficientNetB0 model after training it on the Imagenet dataset [44]. These weights are available from Keras itself. The optimizer used for updating the neural network weights to minimize the cost function used in this work is Nesterov-accelerated Adaptive moment Estimation or Nadam with a learning rate of 0.0001 [45]. Optimizers help us in knowing, how to change weights of the model and learning rate to reduce the occurring loss in training process.

Two callbacks are added before training the model for better accuracy these are, Early Stopping and Reduce Learning Rate on Plateau. Early stopping callback is monitored with the validation loss. If validation loss is not decreasing in 5 continuous epochs, the training is stopped by the callback, and epochs with the least validation loss are saved. This step is taken to prevent the model from overfitting the data and further analyze what can be done to better the model. Reduce Learning rate on the plateau is a callback that also monitors the validation loss and if it does not decrease in a fixed number of epochs which is 3 in this work, reduces the model learning rate by a constant factor F, which is 0.2 in the work.

## V. ANDROID IMPLEMENTATION

This section explains the android implementation of the approach. Any requirement other than an Android device is not necessary according to the implementation that has been used. The whole approach has been ported to the android application using the Android Studio in this work. Further detailed explanations are listed in the given section.

### A. Re-scaling and Grayscale conversion of CT Image

The input image is read in the form of an ARGB (Alpha, Red, Green, Blue) Bitmap, shown in Fig. 8. The uploaded image is checked for its dimensions, which, if not square shape, is rejected and again prompted for input. This bitmap is processed with the Image processing module of the OpenCV library to convert it into a grayscale image with a single channel. This grayscale mat is then resized into a 512 x 512, which is used for further operations

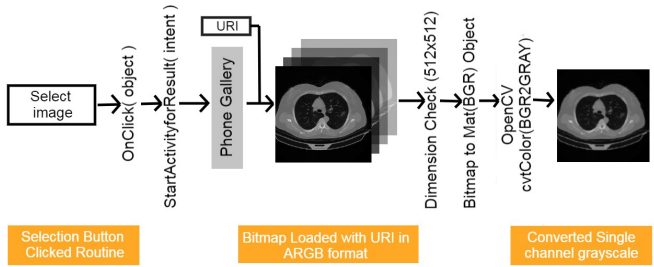

Fig. 8. Figure representing the algorithm - Input of image through user and its pre-processing before moving to inference steps.

The application activity prompts the user to submit the input as a CT image. After uploading is completed, a check is run on the input data to make it in a format acceptable by the further processes, including the segmentation algorithm and the neural network.

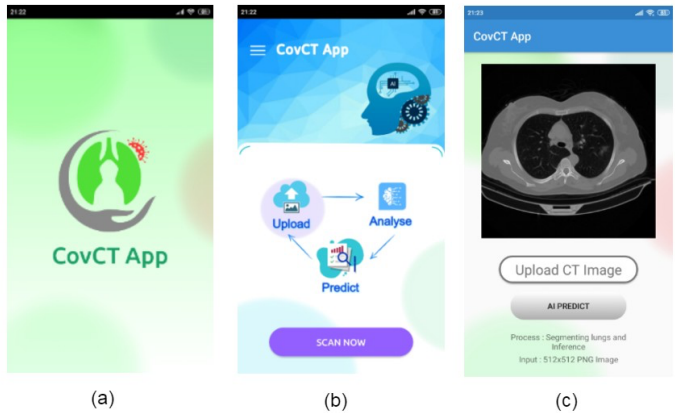

Fig. 9. Image (a) represents the Logo and Splash screen of CovCT Application. (b) Is the Home screen and (c) is the upload activity for accepting inputs from user.

### B. Lung Parenchyma Segmentation

For this stage, the processed input image is passed through the algorithm explained in section IV.C. The image processing module of OpenCV is again put to use. The grayscale single-channel image in the form of a mat object is run through the application. The Image processing algorithm takes a heavy processing, and This algorithm is processed on a worker thread without the main looper in order to prevent it from interrupting or stopping the User Interface ( UI ) Thread.

Message from the worker thread is used to indicate that the algorithm has processed the mat, and upon reception of the message, another routine call is made to the function which handles the inference generation from the original CT image. The process involves using the Imgproc module and core module of OpenCV Software Development Kit (SDK) for implementation into the android application.

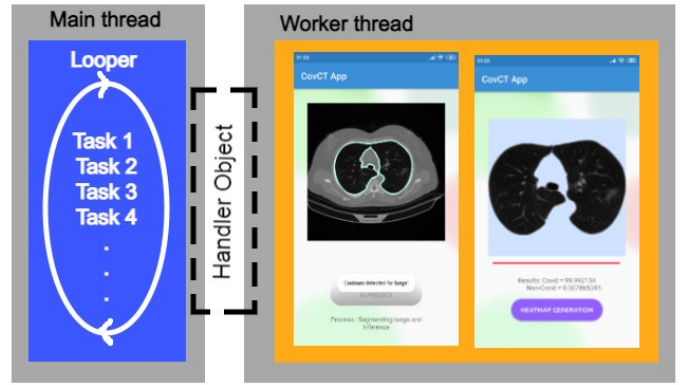

Fig. 10. The process of lung parenchyma segmentation takes place on a separate worker thread. The handler object is use to send and receive message from runnable highlighted in yellow.

### C. Deployment of Neural Network and Inference

The trained model is deployed to the application for offline inference generation instead of relying on the web servers and API programming. As the inference is On-device, it ensures no concern for data privacy, and along with it, fast inference timings are recorded as no data is to be sent or received over the network. For getting the inference from the Neural Network API (NNAPI) of Android OS, the model file is to be bundled into the application. The trained model file is 45 Mb which must be reduced in size to be bundled into the Android Application Package (APK). For this purpose, the model is converted into a Tensorflow lite flat buffer file (.tflite). With the use of Tensorflow lite, not only is the size-reduced, but the model is optimized for speed and latency on the edge devices, as shown in Table IV. The number of threads for the generation of inference are tested on multiple devices for optimal performance in terms of speed and processor efficiency. Starting from 2 threads, the inference timings were reduced till 8 number of threads, but the reduction in timing between 6 and 8 number of threads was insignificant for real-world so, the optimum number of threads for inference generation is chosen to be 6.

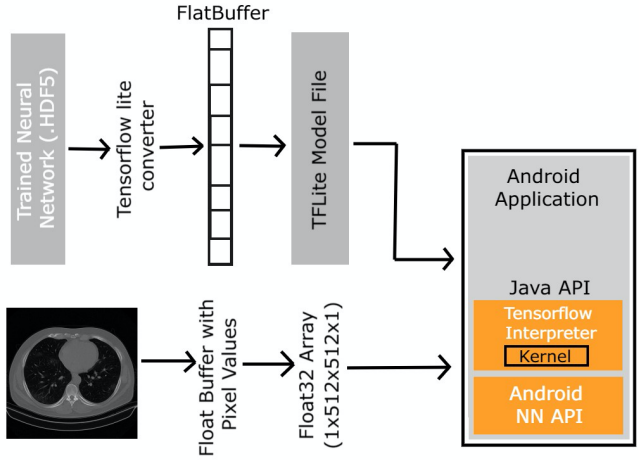

Fig. 11. Graphical representation of deployment and inference generation in android application. The tflite model file is interpreted using Java API of Tensorflow Lite which interacts with the Neural Network API of Android and processes the inputs.

| Device | Inference Time (4 Threads) | Inference Time (6 Threads) | Inference Time (8 Threads) |
|---|---|---|---|
| **Nokia 5.1 Plus** | 291 ms | 273 ms | 276 ms |
| **Xiaomi Note 4** | 303 ms | 270 ms | 270 ms |
| **Samsung A 30** | 276ms | 253 ms | 259 ms |
| **Samsung Galaxy S2** | 312ms | 281 ms | 282 ms |

### D. Heatmap generation using Selective Approach and Multi-Threading

For the generation of the heatmap and saving time and extra computation, gradient-free Class activation Maps (CAM) based visualization method Score-CAM is used [46]. Score-CAM is ported into android using Java in this work. For faster computations, N-Dimensional Arrays for Java are used.

There are 312 activation maps coming as the output of the convolutional layer to be visualized. Each of these activation maps has the size 512 x 512. Processing all of these 312 activation maps by the scorecam algorithm on the android OS takes 600-700 seconds (Based on mobile performance), as shown in Table V. This processing time is too much for the application to be in proper use. To reduce the processing time, a combination of selective approach and multi-threading is used, which significantly reduces the heatmap generation timings, as seen in Table V.

In the selective approach, the number of activation maps was decreased from 320 to 80 by selecting every fourth map starting from the first one, as shown in Fig. 13. This rendered the activation map quickly, compromising a very little clarity of the heat map. Choosing every fourth activation map was proven to give the best and clear results for forming the final heatmap with a reduction of about 74% in heatmap generation timing. Still, the processing time is not suitable for proper usage of the application, as visible from Table V. So multi-threading comes in the role.

| Heatmap Generation Time (Seconds) | Multi-Threading | Selection Approach |
|---|---|---|
| 950s – 1020s | No | No |
| 230s – 250s | No | Yes |
| 370s – 390s | Yes | No |
| **60s – 80s** | Yes | Yes |

Eight worker threads are initialized for processing 10 activation maps, which further reduces the time, as shown in Fig. 12. These threads are run parallel on the android operating system and maximize the use of resources and computation power. The threads are separate from the Main looper, which renders a clean working of the UI thread and keeps the worker thread in the background, reducing about 59% of processing time. Table V shows the results after both the approaches are used together, resulting in a significant reduction of about 93% in heatmap generation process timing.

### E. Augmentation of Heatmap and Gradient Changing

This stage involves the augmentation of the generated heatmap over the grayscale CT image segmented with the help of a binary mask. The generated heatmap is blended with the grayscale CT image. This blending operation takes place according to the equation

$$h(x) = (1 - c) \times f(x) + c \times g(x)$$

[47]

Where, $f(x)$ and $g(x)$ are the source images which are to be blended and $h(x)$ is the final blended image. The $c$ factor is the blending factor in the equation. After the blending is completed, the resultant image of blending operation is masked with help of the binary mask generated in lung segmentation process. This highlights the infected region only on lung parenchyma and is easy to observe by radiologists. For further analysis and highlighting the infected regions using different colors, an option for changing the hue of the resultant image is provided.

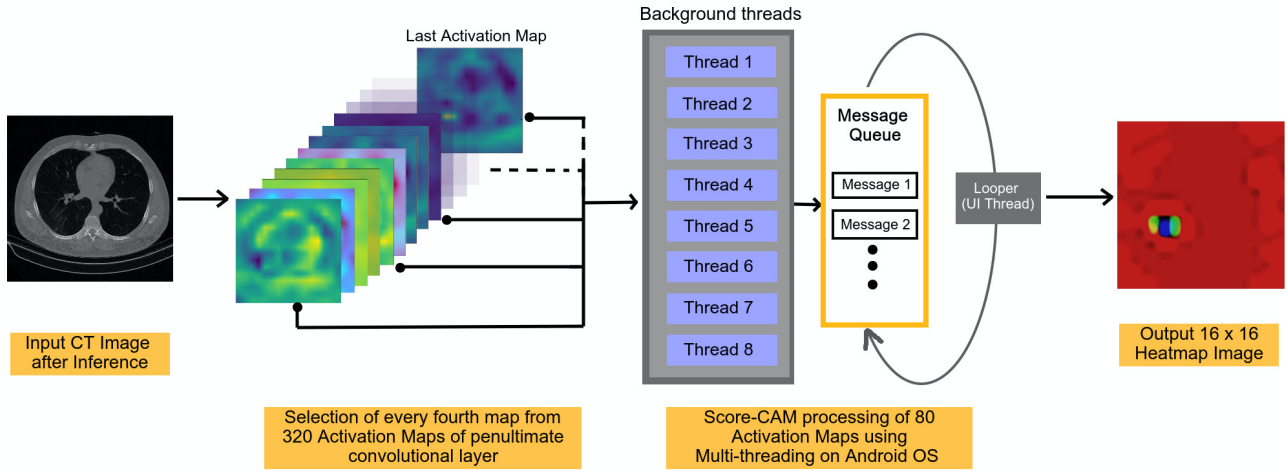

Fig. 12. The representation of Selection and multi-threading approach for generation of Heatmap using ScoreCAM algorithm in Android Device. The background threads keep running without interrupting the Main (UI) thread. The combination of both approaches, reduce the processing time of heatmap significantly and gives an efficient use of CovCT application.

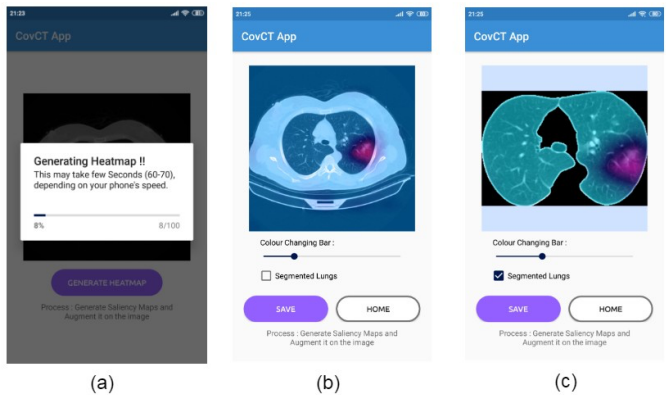

Fig. 13. Figure indicating the last steps of the process of COVID-19 detection from CT scans. (a) image is the processing stage of heatmap generation. (b) is the heatmap augmented over full CT image using linear blending. (c) is the segmented and augmented image using mask.

A track bar is added for changing the hue, which changes the colors in the resultant image and helps to observe the lung in various colors easily. This is done by changing the hue values of the resultant image. The augmented image is converted to the Hue, Saturation, and Value (HSV) color space. The value from the track bar is added to the hue channel by scalar addition, which renders a color changed image. An option to augment the heatmap on the original CT image is also provided, which replaces the masked image with a full CT image as visible in the image (b) of Figure 13.

## VI. RESULTS

Some of the results of our trained model are displayed in Figure 14. These results are generated directly from the python scripts for a swift generation. The results generated from the CovCT android application are also the same (in both position and numerically) but may differ in the color schemes of the detected region in the heatmap. Being of importance in medical diagnosis, our results are also verified by two radiologists, which are discussed in the end of this section.

The training of the model is performed on google collaboratory for an uninterrupted computing and GPU instance. The model is loaded with pre-trained weights on the Imagenet dataset, and transfer learning is done on the training dataset of CT scans. The training accuracy in the first epoch itself went to 79.61%, which increased with each epoch. Fig. 15 shows the graph plotted for the training and validation accuracy of the model in blue and orange color respectively.

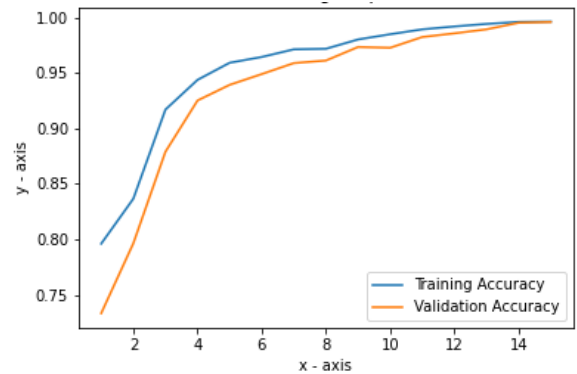

Fig. 15. A graphical representation of training and validation accuracy of the deep learning model after training. The x-axis represents the training epochs and y-axis, the accuracy on a scale of 0 to 1.

It is observed that the training accuracy increased with each epoch and went to 99.62% in the fifteenth epoch. Along with this, it can be observed from this graph that the validation accuracy of the model kept on increasing from 73.34% in the first epoch to 99.58% in the fifteenth epoch, as visible from the graph in Figure 16. Early stopping took place at the fifteenth epoch due to non-decreasing validation loss and to prevent over-fitting.

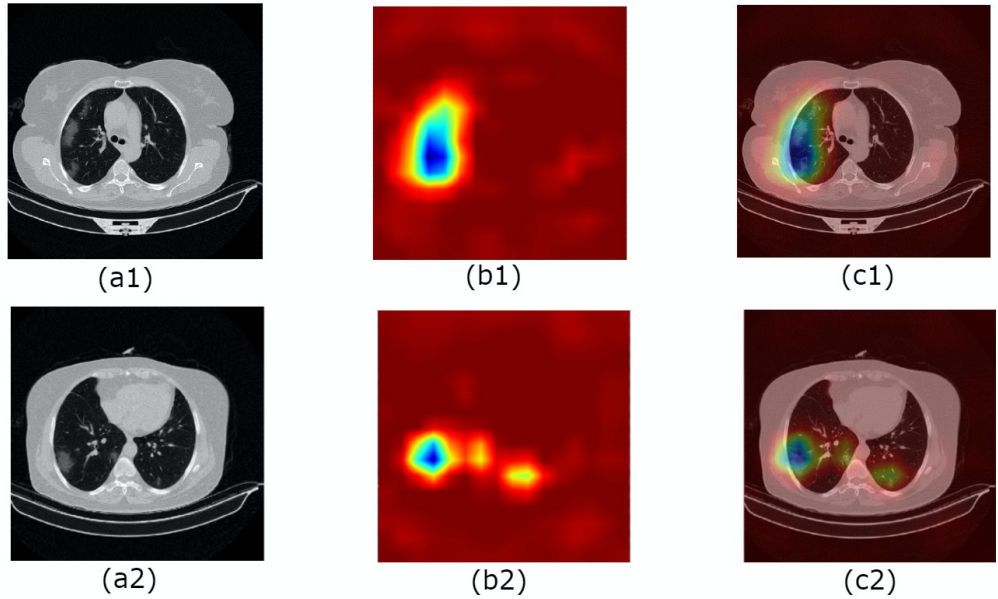

Fig. 14. A sample of data given to radiologists for reviewing the results of our trained model. There are two original CT scans (a1) and (a2), the corresponding heatmaps for detected Covid-19 infection in lung parenchyma is seen in images (b1) and (b2) respectively. Further (c1) and (c2) are their augmented heatmap over original CT scans.

The training phase plots training loss and validation loss using blue and orange lines in Fig. 18, respectively. It is observed that the training loss kept on decreasing till the 15th epoch to 2.11%. The validation loss is decreasing from 28.09% to 2.51%, the minimum at the 15th epoch, after which it does not decrease further and early stopping triggers from preventing overfitting of the model. The observance and the early stopping callbacks saved the best-trained model from the 15th epoch, which has the best validation accuracy of 99.58% and the least validation loss of 2.51%.

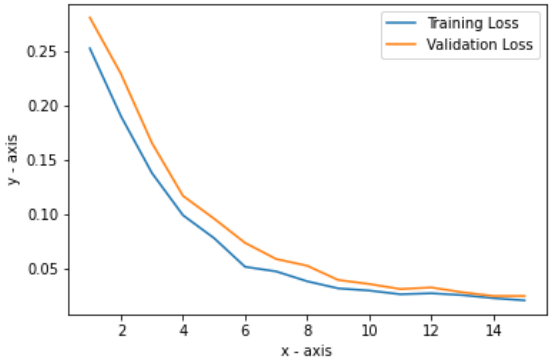

Fig. 16. A graphical representation of training and validation loss of the deep learning model. The x-axis represents the epochs and the x-axis represents the loss on a scale of 0 to 1.

The trained neural network is evaluated on an unseen test partition of the dataset consisting of 4532 images. The test partition consists of 2282 images from patients affected with COVID-19 and 2250 images from normal patients. Taking the COVID-19 CT image as a positive result with class 0 and no-COVID as a negative result with class 1, the model evaluation resulted in 2275 true positives, 12 false positives, 2238 true negatives, and only 7 false negatives as seen through the confusion matrix in Figure 19. Hence, a total of 4513 images are correctly classified by the trained network, and only 19 images were wrong classified.

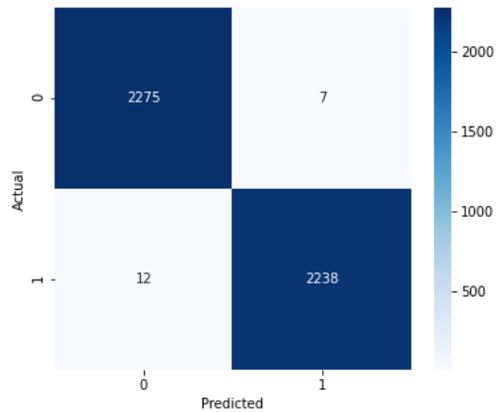

Fig. 17. Confusion Matrix plotted on the result of testing the trained model with the test partition of dataset.

A Receiver Operating Characteristic (ROC) curve is plotted for the trained model. It is a graphic plot that demonstrates the diagnostic ability of a binary classifier system. As our task is binary classification, ROC is plotted, and the respective Area Under Curve (AUC) is shown on the graph itself in Fig. 18. The AUC is the percentage of area under the ROC curve on a scale of 0 to 1. The more the AUC, the more the ability of the model to distinguish between positive and negative classes. Here, in our case the AUC is 0.999843, which shows that our model is highly able to distinguish COVID-19 from normal cases in CT scans and is an excellent classifier.

The precision of the model is calculated to be 99.47% and a sensitivity of 99.69%. The testing accuracy and specificity of the model are 99.58 and 99.46, respectively. The whole summary of testing the model on all five folds of testing unseen data is presented in Table VI. This concludes that the model is very accurate and efficient. Along with this, the model is working efficiently with the PNG images, which do not have features as clear as in TIFF images. The model is ported to an android application and is performing with brilliance on devices with no bugs. The heatmap generation timing range from 60-80 seconds, which is based on the device's performance.

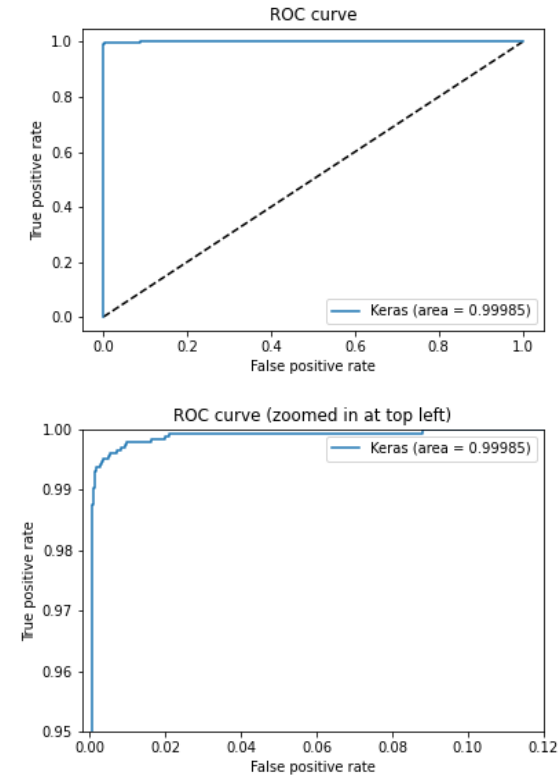

Fig. 18. Top figure represents the ROC curve of the trained model generated with the test data as input. The bottom image is the enlarged view of the top image which shows the in-accuracies in ROC.

### TABLE VI

The Summary of Testing the Model on Five Folds of Unseen Testing Data with Total 4532 Images. The Average row shows the average of result of all Five Folds of Data

| Fold | Performance Metrics (%) | | | | |
|------|----------|-------------|-------------|-----------|-------|
| | Accuracy | Specificity | Sensitivity | Precision | F1 |
| 1 | 99.12 | 98.88 | 99.35 | 98.92 | 99.13 |
| 2 | 99.67 | 99.33 | 100.0 | 99.35 | 99.67 |
| 3 | 99.55 | 99.33 | 99.77 | 99.33 | 99.55 |
| 4 | 99.88 | 99.77 | 100.0 | 99.78 | 99.89 |
| 5 | 99.66 | 100.0 | 99.33 | 100.0 | 99.66 |
| Avg. | 99.58 | 99.46 | 99.69 | 99.47 | 99.58 |

The confidence intervals for our trained classifier are shown in Table VII. Three confidence intervals are calculated at 90%, 95% and 99% along with different number of observations from the validation dataset which include whole dataset, 1000 and 100 observations. The confidence interval at each cell in table VII covers the true classification accuracy of the trained model on validation dataset, which is unseen data to the model.

### TABLE VII

The confidence intervals on various number of observations for Accuracy of the trained model in this work

| Confidence | Number of Observations | | |
|------------|-------------|-------------|-------------|
| | 4532 | 1000 | 100 |
| 90% | 99.43-99.73 | 99.25-99.91 | 98.58-100 |
| 95% | 99.39-99.76 | 99.18-99.98 | 98.32-100 |
| 99% | 99.33-99.82 | 99.05-100 | 97.91-100 |

Two cardiothoracic trained radiologists reviewed the results of our trained model. They found the model to be very sensitive and accurate in detecting ground glass opacities (GGOs). The primary false positives were due to detecting ground glass opacities from non-COVID-19 causes such as pulmonary edema and partially collapsed airspaces. Given the high sensitivity, this could be beneficial in high volume practices and/or resource-poor settings where this application can mitigate delays in diagnosing and triaging.

## VII. Conclusion

The CovCT Application is developed with a very accurate neural network at its base and various techniques applied for its smooth and fast working in this work. The neural network is trained and tested on a vast and balanced dataset of CT images and found to be more accurate and having less number of parameters than other works in the domain.

In figure 19, The performance of the CovCT application is demonstrated on eight different types of lung parenchyma visibility in CT scans from different patients. In (a1) and (a2) Image of Figure 15, both the lung regions are clearly visible in the CT scan, which is processed efficiently and perfectly by the segmentation algorithm as seen in (b1) and (b2) and further (c1) and (c2) are the results of augmentation of heatmap on segmented lungs in which pink shades are representing the COVID-19 affected region. In (a3) image, there is a blur region present in the right lung; still, the segmentation algorithm performs well and segments the lung parenchyma carefully, as seen from image (b3).

In images (a4) and (a5), the right lung is seen with a portion of liver in the CT scan image. Our segmentation algorithm clearly segments the lung from surrounding structures, as seen in results (b4) and (b5) respectively, further the deep learned model also treats that structure as an organ, not opacification, which can be seen in image result (c4) and (c5). In image (a6) one lung is partially visible for the high opacification in lung parenchyma due to parenchyma structures. The prediction of the deep learning model carefully judges the region of covid-19 infection as seen in the resulting image (c6). The structures are not taken as covid-19 related findings. In image (a7), both the lungs are partially visible. Still, they are processed equally well by the segmentation algorithm, as visible from the result (b7).

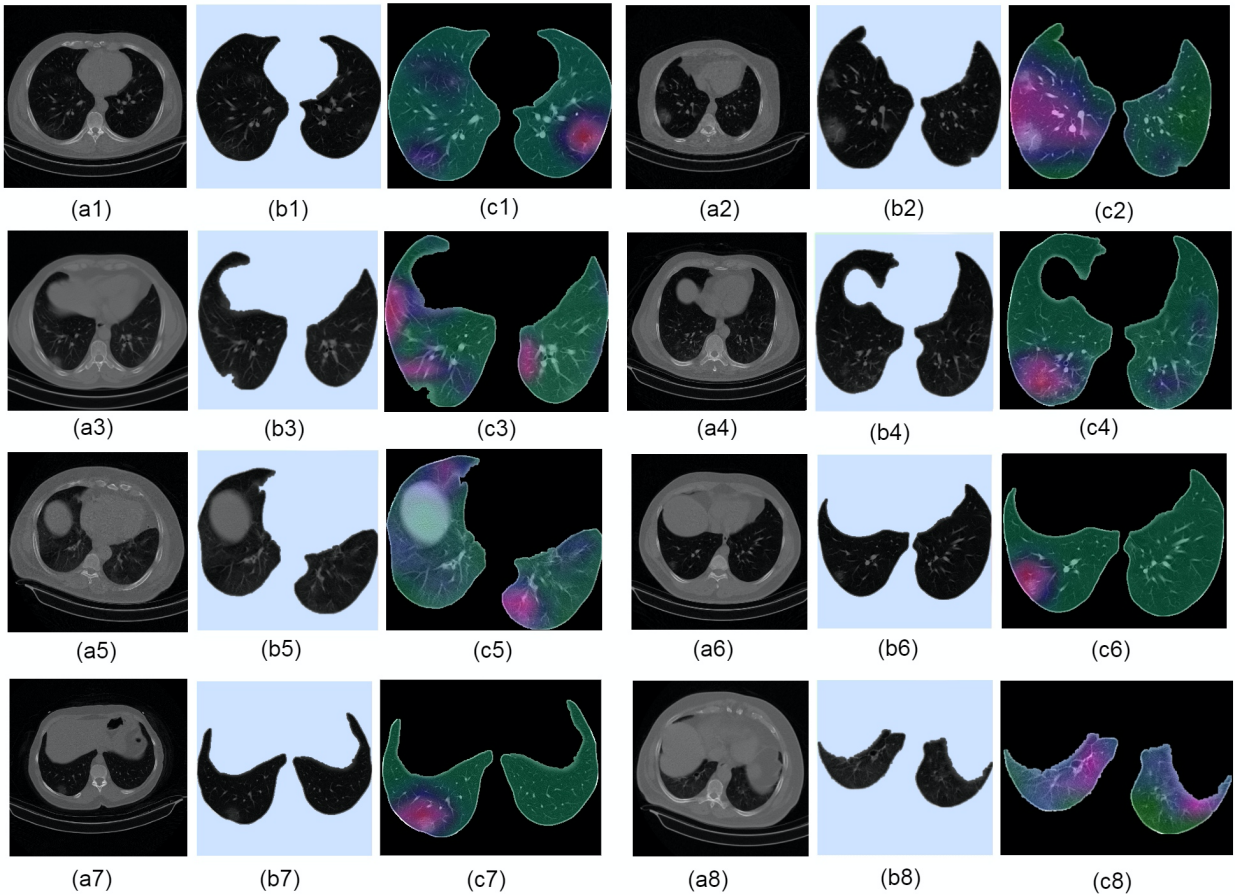

Fig. 19. Analysis of Application's segmentation algorithm and deep neural network of eight typical different lung visibility cases from different patients of COVID-19 where in the first case (a1) and second case (a2), lung parenchyma is fully visible, third case (a3) have a blur in upper part of right lung. Fourth case (a4) shows the lung parenchyma with one lung inflated. Fifth case (a5) is the CT scan with right lung having an organ view in right lung. (a6) image shows right lung inflated, seventh case (a7) shows both lungs partially visible in the image due to inflation. In eight case (a8) again both lungs are inflated and parenchyma is slightly visible. The (b) images are corresponding extracted lungs for each case and (c) images are output with highlighted regions showing COVID-19 related findings using neural network.

Further, the deep learning model can also detect the affected region in this small portion, as seen in image (c7). The test case (a8) is of minimal lung parenchyma visibility and a complex case where the majority of the image contains non-lung structures. The segmentation algorithm does not get confused at any point and segments the parenchyma without mixing it with muscles and fat, as seen in the result (b8). Further, the neural network also judges the opacification and does not get confused due to other structures, visible through the result (c8).

This analysis of the CT scans shows that both, the segmentation algorithm and the neural network perform accurately and efficiently on CT images, including those CT images which do not contain majority lung parenchyma. The application classifies the CT slice into Covid and No-Covid and marks the region of the presence of infection. The features make it an essential utility for early and automated diagnosis of COVID-19.

## VIII. Discussion

This work proposed the development of an AI-driven android application that can detect COVID-19 from chest CT scans by identifying the regions affected by the infection in the segmented lung parenchyma. The app is practical due to its high accuracy, specificity, portability, and easy user interface. There is no need for any exceptional environment created by the user to utilize the application's capabilities. The expert radiologists verified the results of the deep learning algorithm used in application.

While the algorithm may lead to false positive or false negative results, it should serve as a useful tool for healthcare workers to identify patients with COVID-19.

While some other studies have classified Covid-19 from chest CT scans with good accuracy and specificity, results were not verified by a physician or radiologist [48]. Our algorithm has an accuracy of 99.58% and specificity of 99.46% and outlines regions of the Covid-19 presence in the lung parenchyma using heatmaps for easier visualization. Our work is also verified by radiologists. Also in the domain of Covid-19 detection from CT scans using transfer learning methods, our work proves to be more accurate, efficient and of more utility than other works [19][20][21][22][28][34].

Our application is also convenient due to its portable nature. It can reach every part of the globe, helping many doctors and radiologists in the diagnosis of Covid-19. As future research, percent lung involvement can be quantified and help grade patients on severity of disease. Along with this segmentation, better accuracy and mass data adaptation are the areas where more research work can be contributed.

An interesting follow-up project would be to see which of these patients developed progressive lung damage and fibrosis on follow up CT scans. By reviewing this data using AI, we can identify patients at higher risk and alter treatment plans accordingly.

## IX. DATA AND CODE AVAILABILITY

For interested scholars and researchers who wish to use this work for education, research, and improvement purposes, we have maintained an open-source repository containing the codes of the CovCT android application and the deep learning scripts used to train and test the model. The repository contains code files, trained models, deep learning results, and an android application package (APK). A demonstration video is also provided in the repository, along with the android test results. A readme file is embedded into the repository along with the history of whole version control using git. The repository is located at: https://github.com/monjoybme/CovCT_application Along with it a repository for available datasets and resource code files for processing the data is available at: https://github.com/monjoybme/CovCT .

For researchers and scholars who wish to use and get insights from the dataset we used in this work, the authors of the dataset have maintained a well-explained repository having all data and its reports. It can be reached out to here: https://github.com/mr7495/COVID-CTset

## ACKNOWLEDGMENT

Aryan Verma would like to thank Google Summer of Code (GSoC) 2021 program for funding the work and the organization, Department of Biomedical Informatics, Emory School of Medicine for selecting him as a GSoC student and providing constant support, guidance and mentoring for the work. The details of the organization and work done under GSoC can be explored at: https://summerofcode.withgoogle.com/archive/2021/projects/6 468381577838592/

## SUPPLEMENTARY MATERIAL

### I. ARCHITECTURE OF EFFICIENTNETB0

In the transfer learning, we performed for detecting COVID-19 from chest CT scans, the base of the neural network is the EfficientNetB0 architecture without top layers. Choosing this architecture for our task depends majorly on two factors: the fewer parameters and its performance in terms of accuracy when compared against many states of art models. The performance metrics and comparisons have been outlined and displayed in our work. This section explains the reason or exemplary performance by EfficientNet along with fewer parameters.

$$depth: \quad a^x$$
$$width: \quad b^x$$
$$resolution: \quad c^x$$

$$such \ that, \ a.b^2.c^2 \approx 2$$
$$and \quad a \geq 1, b \geq 1, c \geq 1$$

Fig. 1. The formula given by the authors of Efficientnet for compound scaling used for building the model.

While testing various architectures, there are many approaches developed with custom-written architectures by us. A common way in all the approaches to improve the accuracy and performance of the neural network is to scale the model along different dimensions. This scaling may include changing the width of the applied filter, changing the resolution of inputs to various convolutions, and the most used, changing the depth of the model. For example, by increasing the number of layers, ResNet is scaled up from ResNet-18 to ResNet-200, which proves better accuracy. This approach of improving the performance of baselines included manual parameter tuning to find the perfect dimensions, which is not optimal. In order to get a better accuracy and efficiency of the model, it is important to balance all the dimensions of network including its width, depth, and resolution during CNN scaling. This is where the compound scaling method in EfficientNet come into the role.

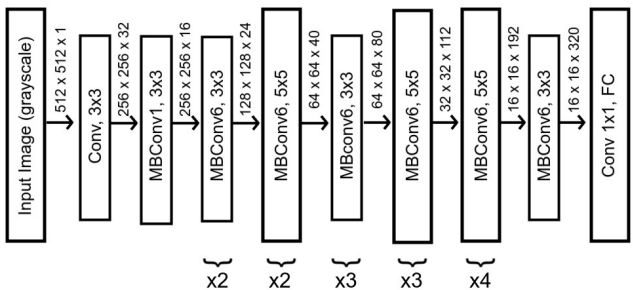

Fig. 2. The architecture of the trained model in the work, with base as EfficientNetB0.

As shown in Figure 1, The method finds the relationship between scaling dimensions, width, depth, and resolution of the baseline network and an appropriate scaling coefficient for each dimension. These coefficients are applied to scale up the baseline network to the desired target model. Here in the above formula, a, b, c distribute the resources to the model's depth, width, and resolution, respectively. The floating-point operations per second (FLOPS) consumed in a convolutional operation are proportional to depth, width$^2$, and resolution$^2$. The authors restrict $a.b^2.c^2$ to 2 so that every new x, the FLOPs needed goes up by $2^x$.

Now, the formula given by the authors in Figure 1 is applicable to scale any CNN architecture, which makes most of the state of art architectures the choice. However, the authors made a new baseline architecture and called it EfficientNet-B0. It is trained with multi-objective neural architecture searches like MnasNet that optimizes both accuracy and FLOPS. This baseline architecture is then scaled to develop a family of EfficientNet models. The architecture of the model is shown in fig 2. The original input size to the EfficientNet models is 224 x 224 x 3, which is not acceptable as our image size is 512 x 512 x 1, so the input node is replaced with another input node.

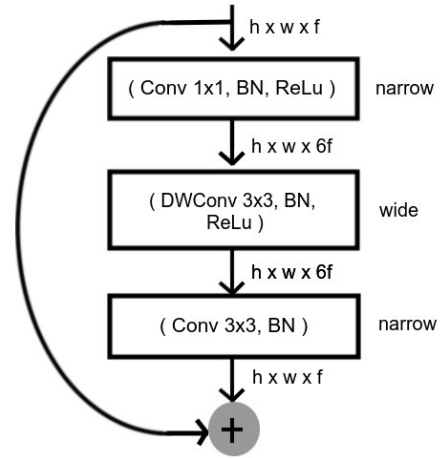

Fig. 3. The graphical representation of a MBConv layer used in the architecture of the Efficientnet family of models. Here h, w, f stands for height, width and number of feature maps in input respectively. BN stands for batch Normalization and DWConv as Depth Wise Convolutional layer.

The base of the neural network is the same as efficientNetB0 till the last convolutional layer. After the single convolution layer in the architecture, the base consists of mobile inverted bottleneck convolution (MBConv) layers, which provide the architecture with a capability to learn efficiently in fewer parameters. MB Conv implements Inverted Residual Blocks, which follows a narrow-broad-narrow approach as shown in Figure 3. The blocks first widen the feature maps with a 1x1 convolution, then use a 3x3 depthwise convolution denoted with DWConv in Figure 3. Lastly, a 1x1 convolution which reduces the number of channels for input and output can be added. This orientation of layers in CNN reduces the number of trainable parameters to a great extent. Finally, The last layers that get stacked after the base of efficientnetb0 are GlobalAveragePooling2D and a dropout layer followed by a softmax to output the results.